\documentclass[aps,pra,twocolumn,showpacs,reprint]{revtex4-1}

\usepackage{bm,bbm,amsmath,bbold,amssymb,amsbsy,txfonts,graphicx,color,hyperref,xfrac}

\newcommand{\kebra}[2]{\vert{#1}\rangle\langle{#2}\vert}
 \newcommand{\tr}[1]{\text{Tr}}
\newcommand{\ket}[1]{|#1\rangle}
\newcommand{\bra}[1]{\langle#1|}

\newcommand{\proj}[1]{\ket{#1}\bra{#1}}

\begin{document}
\title{Signatures of the $A^2$ term in ultrastrongly-coupled oscillators}
\author{Tommaso Tufarelli}
\email{tommaso.tufarelli@gmail.com}
\affiliation{ Blackett Laboratory, Imperial College London, SW7 2BW, UK.}
\author{K. R. McEnery}
\affiliation{ Blackett Laboratory, Imperial College London, SW7 2BW, UK.}
\author{S. A. Maier}
\affiliation{ Blackett Laboratory, Imperial College London, SW7 2BW, UK.}
\author{M. S. Kim}
\affiliation{ Blackett Laboratory, Imperial College London, SW7 2BW, UK.}
\begin{abstract}
We study a bosonic matter excitation coupled to a single-mode cavity field via electric dipole.  Counter-rotating and $A^2$ terms are included in the interaction model, ${\mathbf A}$ being the vector potential of the cavity field. In the ultrastrong coupling regime the vacuum of the bare modes is no longer the ground state of the Hamiltonian and contains a nonzero population of polaritons, the true normal modes of the system. If the parameters of the model satisfy the Thomas-Reiche-Kuhn sum rule, we find that the two polaritons are always equally populated. We show how this prediction could be tested in a quenching experiment, by rapidly switching on the coupling and analyzing the radiation emitted by the cavity. A refinement of the model based on a microscopic minimal coupling Hamiltonian is also provided, and its consequences on our results are characterized analytically.
\pacs{42.50.-p, 42.50.Pq}
\end{abstract}
\maketitle
\section{Introduction}
{\noindent}Quantum technologies exploit intense interactions between field and matter degrees of freedom~\cite{MonroeNATURE}, and it is a typical experimental goal in this context to maximize the coupling between the two. Traditional cavity QED setups have been extremely successful in this regard, yet they result in coupling frequencies that are only a tiny fraction of that of the system components~\cite{HarocheREV}. Experimental advances, for example in semiconductor microcavities and circuit QED, have now pushed the strength of light-matter interactions into the ultrastrong-coupling regime (USC)~\cite{USexp1, USexp2, USexp3, USexp4, USexp5}. This regime is characterized by the coupling frequency $\lambda$ being a non-negligible fraction of the bare frequency of the matter degree of freedom, say $\omega_b$. The theoretical description of the USC goes beyond the rotating wave approximation (RWA), demanding the inclusion in the Hamiltonian of terms that do not conserve the excitation numbers of the individual components --- the `counter-rotating terms' (CR) ~\cite{BraakPRL, CiutiPRA, SolanoPRL}. This regime has been studied extensively due to the lure of exotic phenomena such as the existence of virtual excitations in the ground state~\cite{CiutiPRA}, dynamical Casimir effects~\cite{DeLiberatoCASIMIR}, quantum phase transitions~\cite{CiutiNATURE, BrandesPRA}, and counter-intuitive radiation statistics~\cite{RidolfoPRL,Ridolfo2}. 

{\noindent}In this regime, however, the sole inclusion of the CR terms may not be sufficient to correctly describe the new physics. Another important ingredient is the diamagnetic -- or `$A^2$' -- term, which is proportional to the square of the vector potential $\hat{\mathbf{A}}$ and ensures gauge invariance in the non-relativistic minimal coupling Hamiltonian~\cite{Woolley}. The effects associated with this term, and the related Thomas-Reiche-Kuhn (TRK) sum rules, are of crucial importance in the research on the `Dicke phase transition', and are still under active investigation and debate~\cite{ZakowiczPRL, RzazewskiPRA,Keeling, KnightPRA,BaumannNATURE,CiutiNATURE,ciuticritics,ciutireply,BirulaPRA,ChirolliPRL,DomokosPRL,BambaARXIV}. A further point deserving attention is that the two-level approximation, useful to simplify the description of quantum emitters, may fail in the USC~\cite{threelevel}. Finally,  even the multi-mode nature of the cavity field is known to play a role in the `deep strong coupling' regime $\lambda\gtrsim\omega_b$. \cite{DeLiberatoPRL}.  

{\noindent}In most of the above examples the physics beyond the CR terms, for example due to $A^2$, becomes crucial as the strength of light-matter interactions is pushed towards the extreme regime $\lambda\!\gtrsim\!\omega_b$. In contrast, to the best of our knowledge, clear-cut {\it qualitative} signatures of these extra terms have not been discussed in the currently experimentally relevant regime $\lambda/\omega_b\!\sim\!0.2$. The present work aims at giving a contribution in this direction. We begin by studying a common Hamiltonian model of light-matter interaction, in which a bosonic matter excitation is ultrastrongly coupled to a single-mode cavity field. We find that the $A^2$ term imposes an interesting constraint on the structure of the normal modes of the system, the upper and lower polariton \cite{HopfieldPHYSREV}. This implies that the bare vacuum of the matter and field modes, which is no longer the true ground state of the system, contains equal populations of the two polaritons. Interestingly, this observation is independent of the specific choice of the various model parameters, provided that they are chosen compatibly with the TRK sum rule. To test this finding, one needs to design an experiment that explicitly relies on the relationship between polaritons and bare modes. We show that a possible option is to perform a `quench' of the coupling, a rapid switch-on of $\lambda$ from an initially negligible value, followed by a spectral analysis of the resulting `quantum vacuum radiation' exiting the cavity (that is, radiation that is due to a non-adiabatic change of the ground state of the system) \cite{DeLiberatoCASIMIR}.\\
In the second part of the paper, we investigate the robustness of the considered model by interpreting it as a low-energy approximation to a Coulomb gauge minimal coupling Hamiltonian. This allows us to clarify the role of the TRK sum rule in the considered system, as well as to identify some extra terms -- besides $A^2$ -- that one may need to include in the effective low-energy Hamiltonian to accurately model the USC. In a nutshell, one should include an effective self-interaction term for the matter, mediated by the higher cavity harmonics, plus a term describing the electrostatic interaction between the dipole and its `images' on the cavity walls \cite{DomokosPRL}. While the remarkable symmetry between the two populations is in general lost, we are able to gain an analytical understanding of this more complete model and the consequences of the new terms.

{\noindent}The paper is organized as follows. In section~\ref{basic} we discuss our main result in its simplest form, by analyzing the effect of an $A^2$-like term on a common model of coupled oscillators. Section~\ref{quench} illustrates the quenching experiment that allows to investigate the relationship between bare modes and polaritons. In Section~\ref{micro} we illustrate the microscopic model that is assumed to underlie our theory, clarifying the role of the TRK sum rule and deriving a refined effective Hamiltonian for the two modes of interest. In section~\ref{Dicke} we briefly discuss the extension of our results to Dicke-like models, and in section~\ref{fine} we draw our conclusions. Some additional technical details and derivations are provided in three appendixes.
\section{Basic Model}\label{basic}
{\noindent}We start with a common effective model of light-matter interaction, featuring a photonic mode $\hat a$ of bare frequency $\omega_a$ (cavity mode for brevity) coupled to a bosonic matter mode $\hat b$ of bare frequency $\omega_b$. The latter can be thought of as a quantized oscillating dipole. The Hamiltonian reads ($\hbar\! =\! 1$)
\begin{align}
H &= \omega_a\hat a^{\dagger}\hat a\!+\!\omega_b\hat b^{\dagger}\hat b\!+\!\lambda(\hat a\!+\!\hat a^{\dagger})(\hat b\!+\!\hat b^{\dagger}) \!+\! D(\hat a \!+ \!\hat a^{\dagger})^2\label{Hsys},
\end{align}
where $\lambda$ quantifies the light-matter coupling strength, while the contribution proportional to $D$ is due to the $A^2$ term. As shown in section \ref{micro} below, the TRK sum rule imposes $D\!=\!\lambda^2/\omega_b$~\cite{sumrule}. Nevertheless, we shall keep $D$ implicit for later convenience. Being bilinear in the bosonic operators, Hamiltonian \eqref{Hsys} can be diagonalized exactly. Following Hopfield, we shall refer to the normal modes of the system as the upper (U) and lower (L) polariton \cite{HopfieldPHYSREV}, with associated bosonic operators $\hat p_U$, $\hat p_L$ and eigenfrequencies $\omega_U>\omega_L$. In terms of the normal modes, the Hamiltonian assumes the simple form
\begin{equation}
H=\omega_U\hat p_{U}^\dagger\hat p_U+\omega_L\hat p_L^\dagger\hat p_L,
\end{equation}
up to a constant. Explicit expressions for the eigenfrequencies $\omega_U,\omega_L$ and the polaritonic operators will be shown in the subsection below.
\begin{figure}[t!]
\begin{center}\includegraphics[width=.98\linewidth]{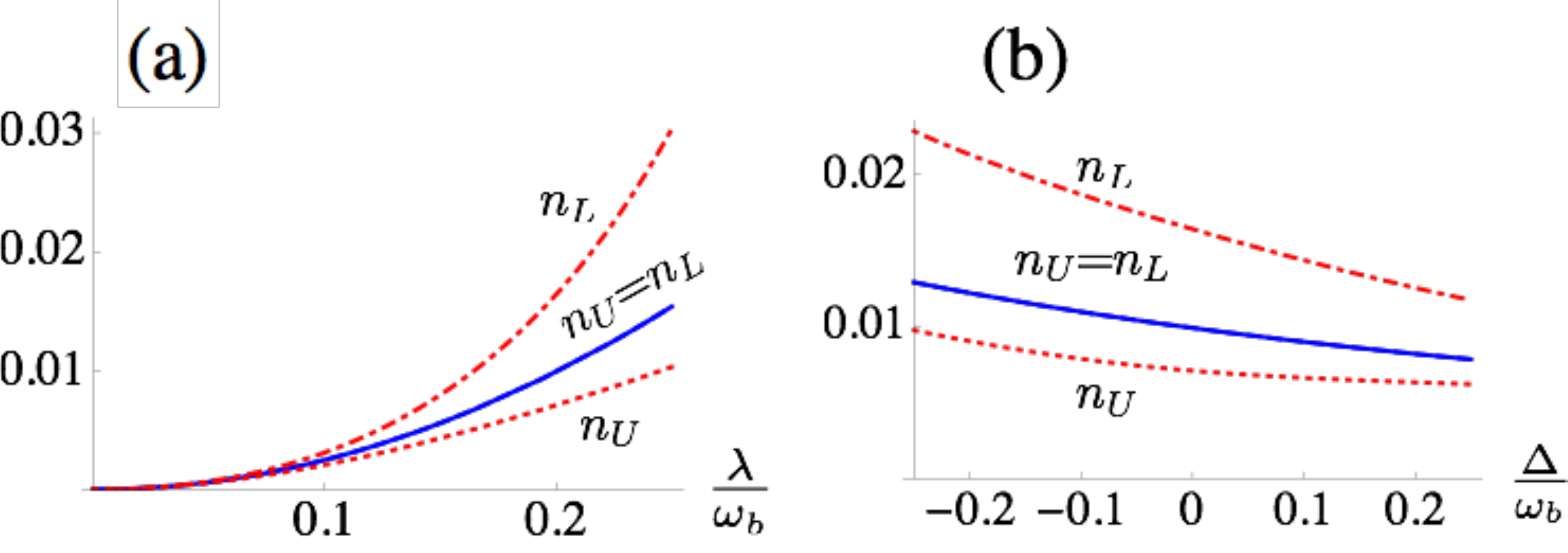}
\end{center}
\caption{(Color online). Mean polaritonic populations $n_U,n_L$ in the bare ground state $\ket 0$. ({\bf a}): Mean populations versus normalized coupling strength at bare mode resonance $\omega_a\!=\!\omega_b$. ({\bf b}):  Mean populations versus bare detuning $\Delta \!=\! \omega_a\!-\!\omega_b$, fixing $\lambda \!=\! 0.2\omega_b$. In all plots, the blue solid line refers to $D\!=\! \lambda^2/\omega_b$, while the red (dotted, dot-dashed) lines to $D\!=\!0$.
\label{fig:Energy}}
\end{figure}
\subsection{A simple signature of the $A^2$ term}\label{signature}
{\noindent}To investigate the impact of $A^2$ on the physics of our system, we shall study in detail the relationship between the bare modes $\hat a,\hat b$ and the polaritons $\hat p_k$, where $k\in\{U,L\}$ hereafter. We start by noting that the bare modes vacuum $\ket0$, defined by $\hat a\ket0=\hat b\ket0=0$, does not in general coincide with the ground state of the polaritons: $\hat p_k\ket0\neq0$. We thus turn our attention to the mean populations 
\begin{equation}
n_k\!\equiv\!\bra0\hat{p}_k^{\dagger}\hat{p}_k\ket0,\label{nk}
\end{equation}
whose nonzero value is perhaps the simplest signature of the USC. Fig.~\ref{fig:Energy} illustrates the behaviour of $n_k$ as a function of the coupling strength $\lambda$, the bare frequency difference $\Delta \!\equiv\!\omega_a\!-\omega_b$ and, most importantly, the parameter $D$. When the TRK value $D \!=\! {\lambda^2}/{\omega_b}$ is taken, we observe that the excitations are distributed equally between $\hat{p}_U$ and $\hat{p}_L$. Setting instead $D\!=\!0$, which corresponds to neglecting $A^2$, predicts a significantly higher population for the lower frequency mode $\hat p_L$. We note that this holds in all the explored range of the remaining parameters $\lambda,\omega_a,\omega_b$. 

{\noindent}This finding can be confirmed analytically. We present here a derivation outlined by an anonymous referee and which is particularly transparent. To find the normal modes of the Hamiltonian, we write $H \!=\!\tfrac12\hat{\boldsymbol{a}}^\dagger\mathbf{H}\hat{\boldsymbol{a}}$ (up to a constant), where $\hat{\boldsymbol{a}}=(\hat a,\hat b,\hat a^\dagger,\hat b^\dagger)$ and $\mathbf{H}$ is a positive matrix that can be easily inferred from Eq.~\eqref{Hsys}. By Williamson's theorem we have  $\mathbf{H}=\mathbf{S}^\dagger{\sf diag}(\omega_U,\omega_L,\omega_U,\omega_L)\mathbf{S}$, where $\mathbf{S}$ is a symplectic matrix~\cite{Williamson}. Hence the polaritonic modes $\hat{\boldsymbol{p}}\!=\!({\hat p}_U,{\hat p}_L,{\hat p}_U^\dagger,{\hat p}_L^\dagger)$ are given by $\hat{\boldsymbol{p}}=\mathbf{S}\hat{\boldsymbol a}\label{polaritons}$, and the bosonic commutation relations $[\hat p_k,\hat p^\dagger_{k'}]=\delta_{kk'}$ are guaranteed by construction. For our specific system, we get (see appendix~\ref{diagonalizza})
\begin{align}
 \!\!\!&\omega_{U,L}^2\!=\!\tfrac{{\omega}_a^2+4D\omega_a+{\omega}_b^2}{2}\!\pm\!\sqrt{\left(\tfrac{{\omega}_a^2+4D\omega_a-{\omega}_b^2}{2}\right)^{\!2}\!+\!4\lambda^{2}\omega_a\omega_b},\label{freqs}\\
	\hat p_{U}\!&=\!\cos\theta[\mu(\!\tfrac{\omega_U}{\omega_a}\!)\hat a\!+\!\nu(\!\tfrac{\omega_U}{\omega_a}\!)\hat a^\dagger]\!-\!\sin\theta[\mu(\!\tfrac{\omega_U}{\omega_b}\!)\hat b\!+\!\nu(\!\tfrac{\omega_U}{\omega_b}\!)\hat b^\dagger],\label{pU}\\
\!\!\hat p_{L}\!&=\!\sin\theta[\mu(\!\tfrac{\omega_L}{\omega_a}\!)\hat a\!+\!\nu(\!\tfrac{\omega_L}{\omega_a}\!)\hat a^\dagger]\!+\!\cos\theta[\mu(\!\tfrac{\omega_L}{\omega_b}\!)\hat b\!+\!\nu(\!\tfrac{\omega_L}{\omega_b}\!)\hat b^\dagger],\label{pL}
\end{align}
where $\mu(x)\!\equiv\!\tfrac12(\sqrt x\!+\!1/\sqrt x)$, $\nu(x)\!\equiv\!\tfrac12(\sqrt x\!-\!1/\sqrt x)$ and $\theta$ is defined by $\cos2\theta\!\equiv\!({{\omega}_a^2\!+\!4D\omega_a\!-\!{\omega}_b^2})/{(\omega_U^2\!-\!\omega_L^2})$; $\theta\!<\!0$. These choices are consistent with the ordering $\omega_U\!>\!\omega_L$. It is easy to check that equations \eqref{pU} and \eqref{pL}, together with their Hermitian conjugates, implicitly define a symplectic matrix ${\mathbf S}$ in accordance with the general discussion above. We can now evaluate $n_k$ by substituting Eqs.~(\ref{pU},\ref{pL}) in Eq.\eqref{nk}, obtaining
\begin{align}
n_U&=\tfrac14\cos^2\theta\,\big(\tfrac{\omega_U}{\omega_a}+\tfrac{\omega_a}{\omega_U}\big)+\tfrac14\sin^2\theta\,\big(\tfrac{\omega_U}{\omega_b}+\tfrac{\omega_b}{\omega_U}\big)-\tfrac12,\label{nU}\\
n_L&=\tfrac14\sin^2\theta\,\big(\tfrac{\omega_L}{\omega_a}+\tfrac{\omega_a}{\omega_L}\big)+\tfrac14\cos^2\theta\,\big(\tfrac{\omega_L}{\omega_b}+\tfrac{\omega_b}{\omega_L}\big)-\tfrac12.\label{nL}
\end{align}
We further notice that the product of the polaritonic frequencies obeys the equation
\begin{align}
{\omega_U\omega_L}\!=\!{\omega_a\omega_b}\sqrt{1\!+\!\tfrac{4}{\omega_a}\big(D\!-\!\tfrac{\lambda^2}{\omega_b}\big)},\label{product}
\end{align}
Choosing the TRK value $D\!=\!\lambda^2/\omega_b$, Eq.~\eqref{product} reduces to $\omega_U\omega_L\!=\!\omega_a\omega_b$, which can be rearranged as ${\omega_U}/{\omega_a}\!\!=\!\!{\omega_b}/{\omega_L}$ and ${\omega_U}/{\omega_b}\!\!=\!\!{\omega_a}/{\omega_L}$. This is easily shown to yield $n_U\!=\!n_L$ via Eqs.~\eqref{nU} and \eqref{nL}. Taking a step further, we find for generic $D\!>\!0$ that the sign of $n_U\!-\!n_L$ is always the same as that of $D\!-\!\lambda^2/\omega_b$ in a broad range of parameters (see appendix~\ref{popstudy}). We thus have a sufficiently general scenario in which the two populations are equal if and only if $D$ assumes the appropriate TRK value, regardless of the specific arrangement of the remaining model parameters. Note that the equality $n_U\!=\!n_L$ can also be stated as a constraint on the matrix elements of ${\mathbf{S}}$, without making reference to a particular quantum state of the system. This simple and yet striking signature of the $A^2$ term on the structure of the polaritons constitutes our main result.

{\noindent}In passing we mention that, in the case of many matter modes interacting with the same single-mode field, a relationship analogue to Eq.~\eqref{product} was derived, implying that the product of polaritonic frequencies equals that of the bare frequencies under the TRK rule \cite{ciuticritics}. It will be interesting to investigate what constraints this may pose to the behaviour of these more general systems.
\section{Detecting $A^2$ via vacuum emission}\label{quench}
{\noindent}In principle, the relationship between bare and polaritonic modes could be investigated via a quenching experiment. The idea is to `switch on' the coupling $\lambda$ and the associated parameter $D$ non-adiabatically, starting from an initially negligible value. If the modulation is applied fast enough, the system remains in its initial state: at sufficiently low temperatures we could assume it to be the bare vacuum $\ket 0$. Since this is no longer the ground state of the Hamiltonian for $t\!>\!0$, the system will relax towards the vacuum of the polaritons, and to do so it must radiate photons outside the cavity. This process is a particular instance of quantum vacuum radiation \cite{DeLiberatoCASIMIR}. In absence of other relaxation mechanisms, we expect $n_k$ photons to be emitted at each frequency $\omega_k$ (on average), so that a simple spectral analysis of the cavity output field could be used to test the equality $n_U\!=\!n_L$ -- see Fig.~\ref{fig:concept}. This intuition is substantiated by the more quantitative discussion below. Before proceeding, we note that the non-adiabatic modulation of light-matter interactions has been experimentally demonstrated in solid state setups, by inducing a fast change in the density of the available charge carriers and hence in the relevant dipole moment matrix elements~\cite{Quench1, Quench2, Quench3}.

{\noindent}To model the radiative relaxation of the system following the quench, we couple the cavity to a continuum of external modes $\hat \alpha_\omega$ -- with $[\hat\alpha_\omega,\hat\alpha_{\omega'}^\dagger]=\delta(\omega\!-\!\omega')$. For simplicity we neglect matter losses  and assume that all modes $\hat \alpha_\omega$ are accessible for measurement. The total Hamiltonian is now
\begin{equation}
H_{\sf tot} =H + H_{\sf ext}+H_I, 
\end{equation}
where $H$ is given by Eq.~\eqref{Hsys}, while $H_{\sf ext}\!=\!\!\int d\omega\,\omega\,\hat \alpha_{\omega}^{\dagger}\hat \alpha_{\omega}$ and $H_I\!=\!\! \int d\omega\, J(\omega)(\hat a+\hat a^{\dagger})(\hat \alpha_{\omega} \!+ \hat \alpha^{\dagger}_{\omega})$ model the free evolution of the external modes and their coupling to the cavity. In the USC, the open dynamics of the system is better described in terms of polaritons. We thus recast $H_{\sf tot}$ in terms of the operators $\hat p_k$, and we assume that the coupling $J(\omega)$ is weak enough for us to perform a RWA in the interaction term $H_{I}$. We obtain
\begin{align}
\!\!H_{\sf tot} \simeq& \!\sum_{k=L, U}\!\!\omega_k\hat p^{\dagger}_k\hat p_k\!+H_{\sf ext}+\!\int\! \!{\rm d}\omega\!\! \sum_{k=L, U}\![J_k(\omega)\hat \alpha^{\dagger}_{\omega}\hat p_k \!+\! {\sf H.c.}]\label{RWA}
\end{align}
where $J_{U}(\omega)\!=\!J(\omega)\!\cos\theta\sqrt{\omega_a/\omega_U}$, $J_{L}(\omega)\!=\!J(\omega)\!\sin\theta\sqrt{\omega_a/\omega_L}$. Note that the RWA must be performed in the polaritonic basis~\cite{Bambareply,BreuerBOOK,ZollerBOOK,Bamba2}, since it is the operators $\hat p_{U},\hat p_L$ that oscillate harmonically in the interaction picture. As we are investigating photon emission in a non-stationary regime, we aim to determine the statistics of the external field modes as a function of the system conditions immediately after the quench. It is convenient to do so by a somewhat unusual application of the Heisenberg equations of motion. We note that the {\it initial} system operators can be expressed as a linear combination of polaritons and external modes {\it at any later time}:  
\begin{figure}[t!]
\centering
\includegraphics[width=\linewidth]{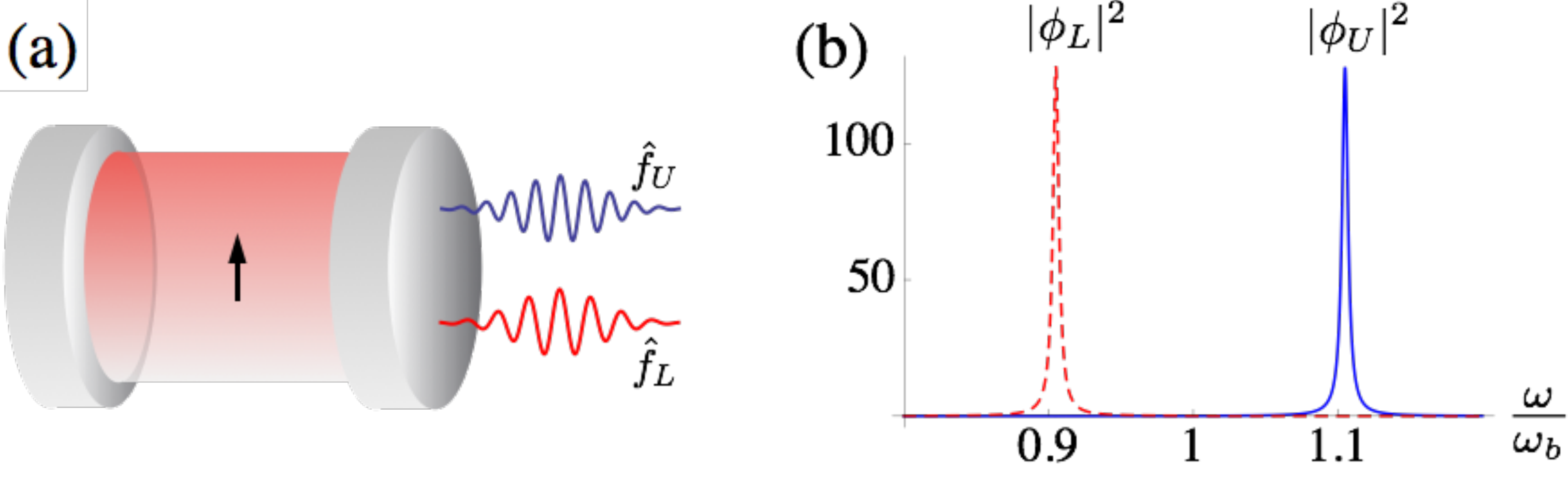}
\caption{(Color online). (a) The coupling between a dipole and a cavity field is suddenly switched on. The system relaxes to the ground state by radiating into the output modes $\hat f_U,\hat f_L$. 
(b) Frequency distribution of the output modes (arbitrary units). We have fixed $\omega_a\!=\!\omega_b$, $\lambda\!=\!0.1\omega_b$, $D\!=\!\lambda^2/\omega_b$ in Eq.~\eqref{Hsys}, and a frequency-independent coupling to the external modes  $J(\omega)\!=\!\sqrt{\gamma/2\pi}$, where $\gamma=0.01\omega_a$ would be the decay rate of the cavity in absence of the matter mode.}
\label{fig:concept}
\end{figure}
\begin{equation}
\hat p_k(0)=\sum_{k'}v_{kk'}(t)\hat p_{k'}(t)+\int{\rm d}\omega\, \phi_k(\omega,t) \hat \alpha_\omega(t),\label{linear}
\end{equation}
{\noindent}where the functions $v_{kk'}(t)$ and $\phi_k(\omega,t)$ 
can be determined as follows. Since the total time derivative must vanish on both sides, the differential equations $\dot v_{kk'}\!=\!i\omega_{k'} v_{kk'}\!+\!i\int{\rm d}\omega\,J_k\phi_k$ and $\partial_t\phi_k\!=\!i\omega\phi_k\!+\!\sum_{k'}J_{k'}v_{kk'}$ must hold, with initial conditions $v_{kk'}(0)\!=\!\delta_{kk'},\phi_k(\omega,0)\!=\!0$. The preservation of commutation relations imposes the normalization $ \sum_l v_{kl}v^*_{k'l}\!+\!\int{\rm d}\omega\,\phi_k\phi_{k'}^*\!=\!\delta_{kk'}$ at all times. For a given form of $J(\omega)$, $v_{kk'}$ and $\phi_k$ could be calculated in principle, e.g. numerically, by Fano-like techniques or Laplace transforms \cite{FanoREV,barnettbook}. Such details, however, are largely unimportant for our purposes. In standard scenarios, Eq.~\eqref{RWA} will induce a dissipative dynamics of the polaritonic system, such that one has ${v}_{kk'}\!\to\!0$ for sufficiently long times, and the full quantum statistics of the initial system state will be retrieved in specific combinations of the external field modes. These can be formally expressed as
\begin{equation}
\hat f_k\equiv\lim_{t\to\infty}\int{\rm d}\omega\, \phi_k(\omega,t) \hat \alpha_\omega(t)=\hat p_k(0).\label{outputmodes}
\end{equation}
By looking at the output modes $\hat f_k$, we can thus access {\it the full quantum statistics} of the polaritons immediately after the quench: the mean populations are for example given by $n_k\!=\!\langle \hat f_k^\dagger\hat f_k\rangle$. We note that the main message expressed by Eq.~\eqref{outputmodes} does not depend on the details of the interaction between cavity and external fields, but each asymptotic amplitude $\tilde\phi_k(\omega)\!\equiv\!\lim_{t\to\infty}\phi_k(\omega,t){\rm e}^{i\omega t}$ does, and needs to be evaluated on a case-by-case basis. Typically, $|\tilde\phi_k(\omega)|^2$ is sharply peaked around the corresponding polaritonic frequency $\omega_k$, and $\hat f_U$ and $\hat f_L$ can be spectrally resolved (an equivalent statement is that the timescales of emission are long as compared to $\omega_k^{-1}$). As an example, in Fig.~\ref{fig:concept} we plot $|\tilde\phi_{k}|^2$ for the simplest case of a frequency-independent coupling to the continuum; we can expect qualitatively similar results when considering more realistic profiles for $J(\omega)$. We remark that the neglect of losses and thermal noise allowed us to derive particularly straightforward relationships between intra- and extra-cavity observables. Still we can expect that, in a realistic system, Eq.~\eqref{outputmodes} can hold to a good approximation if the emission of detectable photons is the dominant relaxation process of the system. A quantitative study of these issues in lossy systems will be presented in future work.
\section{Microscopic model}\label{micro}
{\noindent}In this section we investigate the validity of Hamiltonian \eqref{Hsys} as a low-energy approximation to a more complete microscopic model. This gives us the opportunity to clarify the role of the TRK sum rule in our system, and to discuss some of the possible refinements of our basic model.  In particular, we shall investigate the role of the following contributions to the matter-field interaction: (i) higher harmonics of the cavity field; (ii) the multimode nature of matter excitations; (iii) the electrostatic interaction between the dipole moment of the matter mode and the cavity boundaries. The derivations below are based on the assumption that matter excitations are well described by a collection of quantized harmonic oscillators, in analogy to the Hopfield model \cite{HopfieldPHYSREV}. Our calculations could also be applied to Dicke-like models in the Holstein-Primakoff regime \cite{Holstein,Holstein-multi,CiutiNATURE}, although in that case we are unable to fully take into account the electrostatic dipole-dipole interactions between different atoms. Only the spatially homogeneous contribution of these interactions can be included in our model, by appropriately rescaling the parameter $u$ defined below \cite{Keeling}.
\subsection{The minimal coupling Hamiltonian}
{\noindent}We assume that our matter mode can be microscopically described as a collection of non-relativistic particles of mass $m_j$ and charge $q_j$, subject to a potential $V$ that includes trapping forces as well as inter-particle interactions (in absence of the cavity). The interaction with the electromagnetic field in the cavity is modeled via a minimal coupling Hamiltonian in the Coulomb gauge, as per
\begin{equation}
H_{\sf mic}=\sum_{j}\frac{(\hat{\mathbf{p}}_j-q_j\hat{\mathbf{A}})^2}{2 m_j}+V(\hat{\mathbf{x}}_1,\hat{\mathbf{x}}_2,...)+V_{\sf img}+H_{\sf EM},\label{Hmic}
\end{equation}
where $\hat{\mathbf{p}}_j$ is the momentum of the $j-$th particle, $\hat{\mathbf{x}}_j$ its position, $\hat{\mathbf{A}}$ is the vector potential operator, $V_{\sf img}$ is the electrostatic interaction between matter and cavity walls \cite{DomokosPRL}, and $H_{\sf EM}$ is the free Hamiltonian of the field. We adopt the dipole approximation: the effective linear size of our emitter is assumed to be much smaller than the wavelength of light under consideration, hence the spatial dependence of $\hat{\mathbf{A}}$ across the emitter is neglected. In a similar spirit, we shall assume that $V_{\sf img}$ depends only on the total dipole moment of the matter excitations (in simple geometries, it can thus be calculated with the {\it method of images}). Since the components of $\hat{\mathbf{A}}$ commute with all particle operators, we can expand the Hamiltonian as
\begin{align}
H_{\sf mic}&=H_{\sf mic}^0+H_{\sf int}+H_{\sf EM},\\
H_{\sf mic}^0&=\sum_{j}\frac{\hat{\mathbf{p}}^2_j}{2 m_j}+\hat V,\\
H_{\sf int}&=-\sum_j\frac{q_j\hat{\mathbf{p}}_j\cdot\hat{\mathbf{A}}}{m_j}+\sum_j\frac{q_j^2}{2m_j}\hat{\mathbf{A}}^2+V_{\sf img},
\end{align}
Note that $H_{\sf int}$ includes all the Hamiltonian terms that would be suddenly switched on in the quenching experiment described earlier: indeed, all these terms depend on the effective dipole moment of matter. It is now useful to define 
\begin{align}
\hat{\mathbf{d}}&\equiv\sum_j q_j\hat{\mathbf{x}}_j,\\
\hat{\mathbf{j}}&\!\equiv\!\sum_jq_j\frac{\hat{\mathbf{p}}_j}{m_j},
\end{align}
where $\hat{\mathbf{d}}$ is the {\it electric dipole operator}, while $\hat{\mathbf{j}}$ resembles a current operator (note however that $\hat{\mathbf{p}}_j$ is the canonical momentum, not the kinetic one). We can thus rewrite the microscopic Hamiltonian as
\begin{align}
H_{\sf mic}=H_{\sf mic}^0-\hat{\mathbf{j}}\cdot\hat{\mathbf{A}}+\sum_j\frac{q_j^2}{2m_j}\hat{\mathbf{A}}^2+V_{\sf img}+H_{\sf EM}\label{Hmic1}.
\end{align}
We now recall the TRK sum rule. Let $H_{\sf mic}^0$ have a complete set of eigenstates $\ket{E_n}$ with associated eigenvalues $E_n$ (these would be the eigenstates of matter in absence of interaction with radiation). We recall that the completeness relation $\sum_n\proj{E_n}=\mathbb I$ holds in the Hilbert space of the matter degrees of freedom, and we set the energy of the bare ground state $\ket {E_0}$ to zero for convenience. Exploiting the commutation relations 
\begin{align}
[H_{\sf mic}^0,\hat{\mathbf{d}}]&=-i\hat{\mathbf{j}},\label{Hdj}\\
[\hat d_{k},\hat j_l]&=i\delta_{kl}\sum_jq_j^2/m_j,
\end{align}
we can derive the equality \cite{sumrule}:
\begin{align}
	\sum_n\frac{\bra{E_0}\hat{\mathbf{j}}\cdot\hat{\mathbf{A}}\kebra{E_n}{E_n}\hat{\mathbf{j}}\cdot\hat{\mathbf{A}}\ket{E_0}}{E_n}=\sum_j\frac{q_j^2}{2m_j}\hat{\mathbf{A}}^2. \label{TRK}
\end{align} 
Eq.~\eqref{TRK} is a possible formulation of the TRK sum rule for the ground state. Note that it is an equality for {\it field operators}, since the matrix elements are only taken in the Hilbert space of the matter degrees of freedom. We anticipate that from Eq.~\eqref{TRK} it is possible to derive the crucial equality $D=\lambda^2/\omega_b$ by a somewhat crude two-mode approximation, in which one substitutes $\hat{\mathbf{j}}\simeq-\mathbf{j}_1(\hat b+\hat b^\dagger)$ and $\hat {\mathbf A}\simeq\mathbf A_1(\hat a+\hat a^\dagger)$ ($\mathbf{j}_1$ and $\mathbf A_1$ are constant vectors of the appropriate units -- see below). Consistency with Eq.~\eqref{TRK} then implies ${(\mathbf{j}_1\cdot\mathbf A_1)^2}/{\omega_b}=\mathbf A_1^2\sum_j{q_j^2}/{2m_j}$. Finally, the equality  of interest is obtained if one notices that the relevant coupling constants, in our notation, are given by $\lambda=\mathbf{j}_1\cdot\mathbf A_1$ and $D=\mathbf A_1^2\sum_jq_j^2/2m_j$.

{\noindent}In what follows, we shall show that the above reasoning is indeed correct under certain approximations. Inspired by the Hopfield model \cite{HopfieldPHYSREV}, we now assume that the matter degrees of freedom are well-approximated by a collection of harmonic excitations $\hat b_l$ of frequency $\omega_{b,l}$, and that the relevant matter operators can be expanded as
\begin{align}
H_{\sf mic}^0&\simeq\sum_l\omega_{b,l}\hat b_l^\dagger\hat b_l,\label{H0mic}\\
\hat{\mathbf{j}}&\simeq-\sum_l\mathbf{j}_l(\hat b_l+\hat b_l^\dagger),\label{jmic}\\
\hat{\mathbf{d}}&\simeq-i\sum_l\frac{\mathbf{j}_l}{\omega_{b,l}}(\hat b_l-\hat b_l^\dagger)\label{dmic},
\end{align}
where the constant vectors $\mathbf{j}_l$ encode information about the amplitude and polarization of matter excitations, and we have maintained consistency with Eq.~\eqref{Hdj}. The meaning of the approximation signs in Eqs.~\eqref{H0mic}-\eqref{dmic} is discussed in more detail in appendix~\ref{appb}; the bottom line is that the additional excitations of matter not captured by the modes $\hat b_l$ can be adiabatically eliminated, and their contribution drops out from both sides of the equal sign in the TRK sum rule \eqref{TRK}. We can now expand the vector potential of the field, at the location of the matter mode, as 
\begin{align}
\hat{\mathbf{A}}\!=\!\sum_k\mathbf{A}_k(\hat a_k\!+\!\hat a_k^\dagger), 
\end{align}
where each $\mathbf{A}_k$ is a constant vector characterising the single-photon field amplitude and polarization of the $k$-th cavity mode, with associated bosonic annihilation operator $\hat a_k$. The full Hamiltonian \eqref{Hmic1} can thus be recast as
\begin{align}
H_{\sf mic} & \!=\! \sum_l\omega_{b,l}\hat b_l^\dagger \hat b_l\!+\!\!\sum_{k}\!\omega_{a,k}\hat a_k^\dagger\hat a_k\!+\!\sum_{k,l}\! \lambda_{l,k}(\hat b_l\!+\!\hat b_l^\dagger)(\hat a_k\!+\!\hat a_k^\dagger)\nonumber\\
&+\sum_{k,n}\!D_{kn}(\hat a_k\!+\!\hat a_k^\dagger)(\hat a_n\!+\!\hat a_n^\dagger)+V_{\sf img},
\label{Hsys1}
\end{align}
where $\omega_{a,k}$ is the bare frequency of the $k$-th cavity mode, $\lambda_{lk}=\mathbf{j}_l\cdot\mathbf{A}_k$ quantifies the coupling strength between the $l$-th matter excitation and $k$-th cavity mode, and consistency with the sum rule in Eq.~\eqref{TRK} fixes $D_{kn}\!=\!\sum_l\lambda_{l,k}\lambda_{l,n}/\omega_{b,l}$ ~\cite{sumrule}. Note that we have not specified yet the electrostatic contribution $V_{\sf img}$: here we shall not attempt to study the structure of this term from first principles, rather  we will assume that it is a quadratic function of the dipole moment $\hat{\mathbf{d}}$; this corresponds to the assumption that the induced charge densities on the cavity walls will depend linearly on the dipole moment components.
\subsection{Reduction to a two-mode model}
{\noindent}To recover a simpler Hamiltonian resembling Eq.~\eqref{Hsys}, we shall now assume that the lowest-frequency modes $\hat a\!\equiv\!\hat a_1$ and $\hat b\!\equiv\!\hat b_1$ are dominant in the interaction. Intuitively, this should hold when the two frequencies $\omega_a\!\equiv\!\omega_{a,1}$ and $\omega_b\!\equiv\!\omega_{b,1}$ are of the same order, the coupling $\lambda\!\equiv\!\lambda_{1,1}$ is not too large (as compared to $\omega_a,\omega_b$), and the TRK sum rule is approximately saturated by the considered transition: $D_{1,1}\!\sim\!\lambda^2/\omega_{b}$ (this also implies that $\mathbf{j}_1$ and $\mathbf{A}_1$ should be approximately parallel to each other). All the remaining parameters should conspire in such a way that the other light and matter modes will either stay close to their ground state in the dynamics of interest, or they will decouple from $\hat a,\hat b$ (e.g. by featuring polarizations orthogonal to both $\mathbf{j}_1$ and $\mathbf{A}_1$). Under these conditions we can obtain an effective Hamiltonian for the two modes $\hat a,\hat b$ by adiabatically eliminating all other modes from Eq.~\eqref{Hsys1}. Following standard procedures to eliminate weakly coupled excitations (See Refs.~\cite{james,sorensen} and appendix~\ref{appb}) we thus obtain the effective Hamiltonian
\begin{align}
H_{\sf eff}&= \omega_a\hat a^{\dagger}\hat a\!+\!\omega_b\hat b^{\dagger}\hat b\!+\!\lambda(\hat a\!+\!\hat a^{\dagger})(\hat b\!+\!\hat b^{\dagger}) \!+\! D(\hat a \!+ \!\hat a^{\dagger})^2\nonumber\\
&-\eta(\hat b\!+\!\hat b^\dagger)^2+u[i(\hat b\!-\!\hat b^\dagger)]^2
\label{adiabatic}
\end{align}
where $\eta\!\equiv\!\sum_{k>1}{\lambda_{1,k}^2}/{\omega_{a,k}}$, $u$ is a phenomenological coupling parameter arising from $V_{\sf img}$ (see below), and we have kept only second order terms in the couplings $\lambda_{l,k}$ with $l>1$. We observe that the elimination of the off-resonant matter modes induces the rescaling $D_{1,1}\!\to\! D_{1,1}\!-\!\sum_{l>1}\lambda_{l,1}^2/\omega_{b,l}$, hence retrieving the same value $D$ that we used in Hamiltonian \eqref{Hsys}. This suggests that one should {\it not} include the contributions of neglected transitions in the $A^2$ term, and justifies our slight abuse in terminology in referring to $D=\lambda^2/\omega_b$ as the `TRK sum rule'. The terms proportional to $\eta$ and $u$ are qualitatively new contributions that were not present in Eq.~\eqref{Hsys}. While the coefficient $\eta$ depends on the specific cavity structure, it is in general positive and of second order in the coupling, such that it may not be negligible with respect to the other terms. We emphasize that the $\eta$-term has {\it not} been obtained through a canonical transformation of $H_{\sf mic}$, even though its form may be reminiscent of the ``$P^2$ term" arising in the Power-Zienau-Woolley representation \cite{Cohen}. Finally, the term proportional to $u$ is simply the contribution of the matter mode $\hat b$ to the electrostatic energy $V_{\sf img}$ (recall that we are assuming $V_{\sf img}$ to be quadratic in the dipole moment -- see also appendix~\ref{appb}). Since we are not aware of a general method to determine the parameter $u$ as a function of the others, as we did for example with $D$, we shall study its effect as it is varied in the range $u\in[-D,D]$.
\subsection{Reliability of the effective Hamiltonian}
\begin{figure}[t!]
\begin{center}
\includegraphics[width=\linewidth]{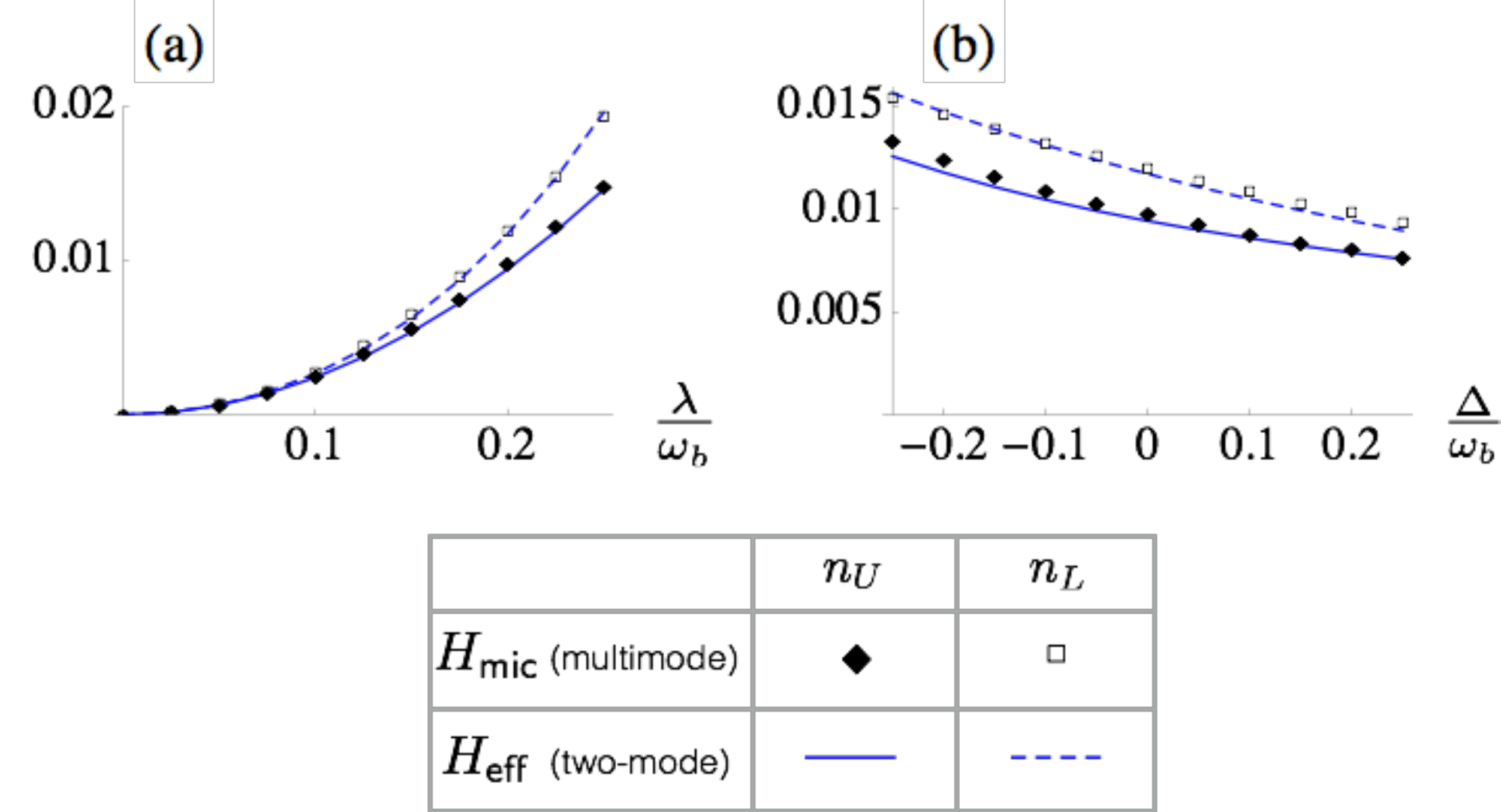}
\end{center}
\caption{(Color online). Comparison between the polaritonic populations predicted by the multimode Hamiltonian ~\eqref{Hsys1} and the effective model \eqref{adiabatic}, in the case $u\!=\!0$. {\bf (a)}:  Populations versus coupling strength, when $\omega_a\!=\!\omega_b$. {\bf (b)}: Populations versus bare detuning $\Delta=\omega_a-\omega_b$, with $\lambda\!=\!0.2\omega_b$. To perform the simulations it was sufficient to include 5 matter transitions and 25 cavity modes in Hamiltonian \eqref{Hsys1}.} 
\label{fig:More}
\end{figure}

{\noindent}To confirm the validity of the effective Hamiltonian \eqref{adiabatic}, we compare its predictions to those of Eq.~\eqref{Hsys1} in a concrete example. For definiteness we assume that all the vectors $\mathbf{j}_l$ and $\mathbf{A}_k$ characterizing the modes of interest lie along the same axis, and we fix the structure of cavity modes and matter excitations such that $\omega_{a,k}\!=\!(2k\!-\!1)\omega_a$, $\omega_{b,j}\!=\!\tfrac13(4j^2\!-\!1)\omega_b$ and $\lambda_{j,k}\!=\!\lambda \tfrac{3j}{(4j^2\!-\!1)\sqrt{2k\!-\!1}}$, mimicking the relevant frequencies and coupling constants for a deep rectangular well placed in the middle of a Fabry-Perot resonator (with the important difference that, for us, each matter excitation is associated with a different harmonic oscillator). As a result we obtain $\eta\simeq0.23\lambda^2/\omega_a$. For the purposes of this section we simply take $u=0$, as our primary objective is to show that the introduction of the $\eta$ term and the rescaling of the $A^2$ term capture well the effect of higher-frequency cavity modes and matter excitations.
Fig.~\ref{fig:More} displays a comparison of the populations $n_U,n_L$ in the bare ground state, as predicted by either $H_{\sf mic}$ or $H_{\sf eff}$. In the former case, the lower (upper) polariton can be defined as the eigenmode of $H_{\sf mic}$ with the lowest (second lowest) frequency. In both cases we can observe a small deviation from the results of Fig.~\ref{fig:Energy}, such that $n_L\!\gtrsim\!n_U$. The important point is that the impact of the fuller matter-field interaction model is well captured by the simple effective Hamiltonian ~\eqref{adiabatic}: the discrepancy between Eqs.~(\ref{Hsys1}) and (\ref{adiabatic}) ranges from $1\%$ to $7\%$ of the plotted quantities. Interestingly, the best agreement is observed when $\omega_a\!\sim\!\omega_b$ and $\lambda/\omega_b\sim\!0.25$.
\subsection{Effective Hamiltonian analysis: distribution of populations}
\begin{figure}[t!]
\begin{center}
\hspace{-.5cm}\includegraphics[width=\linewidth]{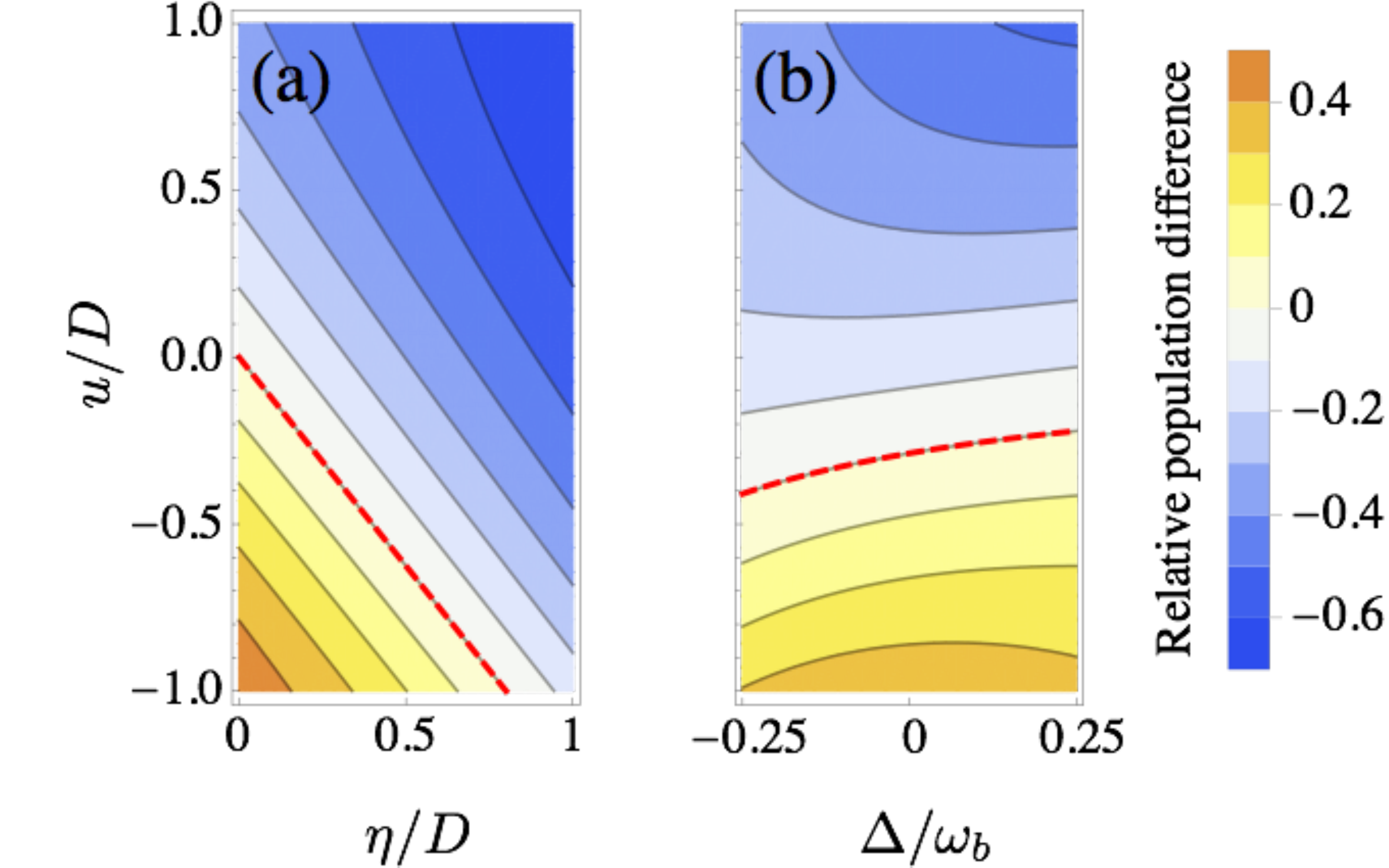}
\end{center}
\caption{(Color online). Contour plots of the relative population difference $(n_U\!-\!n_L)/(n_U\!+\!n_L)$ predicted via the few-mode Hamiltonian $H_{\sf eff}$. We take $D=\lambda^2/\omega_b,\lambda=0.25\omega_b$ in all plots. In plot {\bf (a)} we fix $\omega_a=\omega_b$ and vary the two parameters $u,\eta$, while in plot {\bf (b)} we fix $\eta=0.23\lambda^2/\omega_a$ (as obtained earlier for the Fabry-Perot modes) and vary $u$ together with the bare-mode detuning $\Delta$. The red dashed line corresponds to the model parameters obeying Eq.~\eqref{equalpops}, resulting in $n_U\!=\!n_L$. We have obtained qualitatively similar plots for $\lambda\in\{0.2\omega_b,0.15\omega_b,0.1\omega_b,0.05\omega_b\}$ (Not shown).\label{fig:Heff}} 
\end{figure}

{\noindent}One of the advantages of a few-mode model is that it can be amenable to analytical investigations. Having provided some evidence for the reliability of the Hamiltonian $H_{\sf eff}$, here we exploit its relatively simple form to generalize the results of Section \ref{signature}, and discuss how the parameters $\eta,u$ influence the balance of polaritonic populations in the bare ground state. As we will see shortly, one can still determine a simple analytical condition on the model parameters that results in equal populations. Following similar steps as in Section~\ref{signature}, it is possible to obtain explicit expressions for the eigenfrequencies $\omega_{U,L}$ and the corresponding polaritonic operators. For brevity we shall report the expressions of the bare ground state populations, referring the reader to appendix~\ref{diagonalizza} for a full diagonalization of the Hamiltonian. The quantities of interest read
\begin{align}
n_U&=\tfrac14\cos^2\theta\,\big(\tfrac{\omega_U}{\omega_a}\!+\!\tfrac{\omega_a}{\omega_U}\big)\!+\!\tfrac14\sin^2\theta\,\big(\tfrac{\omega_U}{\omega_b+4u}\!+\!\tfrac{\omega_b+4u}{\omega_U}\big)-\tfrac12,\label{nU1}\\
n_L&=\tfrac14\sin^2\theta\,\big(\tfrac{\omega_L}{\omega_a}\!+\!\tfrac{\omega_a}{\omega_L}\big)\!+\!\tfrac14\cos^2\theta\,\big(\tfrac{\omega_L}{\omega_b+4u}\!+\!\tfrac{\omega_b+4u}{\omega_L}\big)-\tfrac12.\label{nL1}
\end{align}
The specific forms of $\theta,\omega_U,\omega_L$ do not enter the current discussion, but it is important to point out that they will be different from what reported in Section \ref{signature}, except for the `trivial' case $\eta=u=0$. Of great use to us is the following product rule obeyed by the polaritonic frequencies:
\begin{align}
\omega_U^2\omega_L^2=\omega_a(\omega_b\!+\!4u)[(\omega_a\!+\!4D)(\omega_b\!-\!4\eta)-4\lambda^2]\label{prod2},
\end{align}
which can be exploited as follows. By inspecting Eqs.~\eqref{nU1} and \eqref{nL1} we see that a {\it sufficient} condition to achieve equal populations is now given by $\omega_U\omega_L=\omega_a(\omega_b\!+\!4u)$. Comparing this with Eq.~\eqref{prod2}
we can then derive the following condition:
\begin{align}
u=-\frac{\omega_a+4D}{\omega_a}\eta+\frac{\omega_b}{\omega_a}\left(D-\frac{\lambda^2}{\omega_b}\right)\quad\Rightarrow\quad n_U\!=\!n_L\label{equalpops}
\end{align}
Obviously, $D=\lambda^2/\omega_b,\eta=u=0$ represents a possible solution, which corresponds to what we found in Section~\ref{signature}. In general, we can see that assigning the TRK value to the parameter $D$ is no longer sufficient to ensure equal populations. As shown in Fig~\ref{fig:Heff}, the distribution of populations will be ultimately determined by the additional model parameters. It is interesting to note that, while Eq.~\eqref{equalpops} is only a sufficient condition to have  $n_U\!=\!n_L$, it is both necessary and sufficient in the examples reported in Fig.~\ref{fig:Heff}, where we allow $\eta$ and $u$ to be of the same order as the parameter $D$. In a rather broad range of parameters, we thus have a simple analytical criterion to determine which polariton will be most populated.
\section{Extension to Dicke models}\label{Dicke}
\begin{figure}[t!]
\begin{center}
\includegraphics[width=\linewidth]{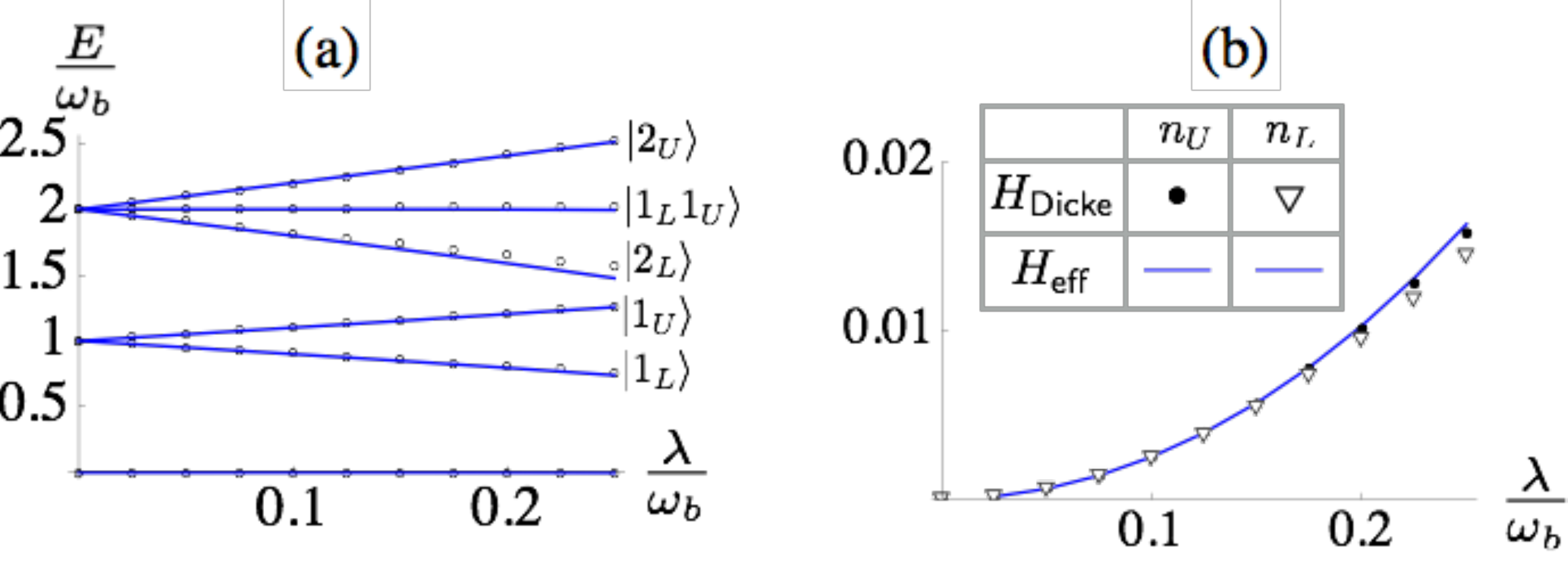}
\end{center}
\caption{(Color online) {\bf (a)}: Low-energy spectra of $H_{\sf Dicke}$ with $n=5$ (empty black circles) and $H_{\sf eff}$ (blue, continuous) as a function of coupling strength. We have displayed the rescaled energies $E/\omega_b$ of the ground state and the lowest five excited states (see labels in the plot). We fixed $\omega_a=\omega_b$, $D=\lambda^2/\omega_b$, $\eta=0.23\lambda^2/\omega_a$. The parameter $u$ is chosen to satisfy Eq.~\eqref{equalpops}, such that equal populations are predicted by $H_{\sf eff}$. {\bf (b)}: Comparison of the polaritonic populations in the bare ground state, for the two Hamiltonias $H_{\sf eff}$ and $H_{\sf Dicke}$. The discrepancy between the two increases with the coupling strength $\lambda$.} 
\label{fig:HDicke}
\end{figure}

{\noindent}Before concluding, it is useful to illustrate the modification of our predictions when the behaviour of matter deviates significantly from a simple harmonic oscillator. To this end, we consider a generalized Dicke model that closely mimics Eq.~\eqref{adiabatic}: 
\begin{align}
\!\!\!H_{\sf Dicke}&=\omega_a\hat a^\dagger\hat a+\omega_b \frac{\hat J_z}{2}+\lambda\frac{\hat J_x}{\sqrt{n}}(\hat a\!+\!\hat a^\dagger)\!+\!D(\hat a\!+\!\hat a^\dagger)^2\nonumber\\
&-\eta\frac{(2\hat J_x)^2}{n}+u\frac{(2\hat J_y)^2}{n},\label{angolone}
\end{align}
where $\hat J_x,\hat J_y,\hat J_z$ are spin-$n/2$ operators. Note that, through the Holstein-Primakoff mapping, it is possible to recover the Hamiltonian $H_{\sf eff}$ as the limit of Eq.~\eqref{angolone} for $n\to\infty$ \cite{Holstein,Holstein-multi}. In Dicke models the integer $n$ is typically interpreted as the number of two-level atoms that collectively interact with the same field. However, as discussed in section~\ref{micro}, in this case the Hamiltonian $H_{\sf Dicke}$ does not fully take into account the impact of electrostatic dipole-dipole interactions, an approximation that might not be well justified in the ultrastrong coupling regime \cite{DomokosPRL}.  While these issues certainly deserve further study, here we shall simply adopt Eq.~\eqref{angolone} as our starting point. For our scopes the integer $n$ quantifies the importance of anharmonic effects in matter, such that $H_{\sf Dicke}$ can be interpreted as $H_{\sf eff}$ plus a anharmonic perturbation (this could be formalized via the Holstein-Primakoff mapping, if desired). Exploiting this interpretation we shall draw a comparison between the two models, making use of concepts that are rigorously defined only for the bilinear Hamiltonian $H_{\sf eff}$. We diagonalize Eq.~\eqref{angolone} numerically by truncating the Hilbert space of the cavity, and for our scopes it is sufficient to represent $\hat a$ and $\hat a^\dagger$ as 10-dimensional matrices. Fig.~\ref{fig:HDicke} compares the low-energy spectrum of $H_{\sf Dicke}$, with $n=5$, with that of $H_{\sf eff}$. The qualitative agreement between the two encourages us to label the ground and excited states respectively as $\ket G$, $\ket{N_L,\!N_U}$ in both models. Note that the definition $\ket{N_L,\!N_U}\!\propto\!(\hat p_L^\dagger)^{N_L}\!(\hat p_U^\dagger)^{N_U}\!\ket{G}$ holds in the case of $H_{\sf eff}$, while no simple explicit expression is available for the eigenstates of $H_{\sf Dicke}$. Both Hamiltonians commute with the parity operator: it can be directly verified that each term in either Eq.~\eqref{adiabatic} or Eq.~\eqref{angolone} can only leave the number of bare excitations unchanged, raise it by two or lower it by two. In both models, it can be shown that the ground state $\ket{G}$ is even, while the parity of the excited states is $(N_U\!+\!N_L)\,{\sf mod}\,2$. This symmetry implies that we can expand the bare ground state as
\begin{align}
\ket0\simeq c_0\ket{G}+c_{2_U}\ket{2_U}+c_{2_L}\ket{2_L}+c_{1_L1_U}\ket{1_L1_U}\label{groundo},
\end{align}
where the $c$'s are appropriate complex coefficients, and the overlaps with higher excited states are found to be negligible in all the explored examples. It is understood that the various coefficients and states appearing in Eq.~\eqref{groundo} assume different forms depending on whether $H_{\sf eff}$ or $H_{\sf Dicke}$ is being considered. From Eq.~\eqref{groundo} it follows that the polaritonic populations in the bare ground state are well approximated by
\begin{align}
	n_U&\simeq2|c_{2_U}|^2+|c_{1_L1_U}|^2,\label{nU5}\\
	n_L&\simeq2|c_{2_L}|^2+|c_{1_L1_U}|^2.\label{nL5}
\end{align}
While of no particular use in the study of $H_{\sf eff}$, where exact expressions are readily available, we can exploit Eqs.~\eqref{nU5} and \eqref{nL5} to calculate (and in fact, {\it define}) the populations $n_U,n_L$ for the Dicke model. Fig.~\ref{fig:HDicke} displays the behaviour of the populations of interest for an arrangement of parameters satisfying Eq.~\eqref{equalpops}. A good qualitative agreement can be observed between the two models in the range of coupling strengths $\lambda\lesssim0.25\omega_b$. As it can be expected the discrepancy between the two tends to grow with increasing $\lambda$: differently from $H_{\sf eff}$, which predicts equal populations, $H_{\sf Dicke}$ results in $n_U\gtrsim n_L$. A detailed explanation of this result goes beyond the scopes of this manuscript. In future studies of the ultrastrong coupling regime, it will be certainly interesting to delve deeper in the study of similarities and differences between truly nonlinear Hamiltonians, such as $H_{\sf Dicke}$, and bilinear interaction models such as those studied here.
\vspace{1cm}
\section{Discussion and Conclusions}\label{fine}
{\noindent}We have identified a qualitative signature of the $A^2$ term in what is arguably the simplest model of ultrastrong coupling between a single-mode field and a matter excitation. Our finding is a consequence of the TRK sum rule, and in terms of bare vacuum populations it can be expressed in the elegant form $n_U\!=\!n_L$. We have shown how this prediction can be verified by a quenching experiment, assuming that the dominant decay mechanism of the system is the emission of detectable photons. Taking one step further, we have then questioned the validity of the model itself, by interpreting it as a low-energy approximation to a multimode minimal-coupling Hamiltonian. Our analysis gives rise to an effective Hamiltonian for the two modes of interest, featuring two extra terms as compared to our initial interaction model. The effect of these new terms on the quantities $n_U,n_L$ has been discussed. In fact, the information provided in this manuscript makes it straightforward to characterize the full {\it covariance matrix} of the polaritonic modes in the bare ground state. 

{\noindent}The results of our study are relevant to a regime of ultrastrong coupling that is accessible in state-of-the-art experiments, and can be used to check the validity of various common approximations and assumptions in the interaction model. For example, if the relationship between bare and polaritonic modes could be experimentally investigated (e.g. via the quenching experiment described here), one would be able to estimate the most appropriate values of the various coupling constants appearing in the effective model $H_{\sf eff}$ (obviously, also the two-mode assumption should be verified in parallel). This could be particularly valuable in systems such as circuit QED, where the influence of the TRK sum rule on the model parameters is under debate (for example, it has been suggested that $D\ll\lambda^2/\omega_b$ could be obtained \cite{CiutiNATURE}).

{\noindent}A rather broad and intriguing open question is whether simple signatures of $A^2$ and other Hamiltonian terms can be identified in more general models, for example featuring a larger number of matter and field modes and/or strong anharmonicities. Despite the theoretical challenge, including one or more of these generalizations may become necessary in attempting to model ever increasing light-matter couplings. 
\subsection*{Acknowledgements}
{\noindent}This work was supported by the Leverhulme Trust, the Qatar National Research Fund (Grant No. NPRP 4-554-1-084), the UK EPSRC (Active Plasmonics Programme and Grant No. EP/K034480/1). S.A.M. and M.S.K. acknowledge support from the Royal Society Wolfson Research Merit Awards. We thank F. Armata, S. Barnett, F. Ciccarello, G. M. Palma, J. Iles-Smith, M.-J. Hwang, R. Passante, P. L. Knight, and S. De Liberato for fruitful discussions.

\noindent\textit{Note added} ---  During the revision of this manuscript, a preprint was published dealing with the detection of the $A^2$ term in circuit QED \cite{Apero}.
\newline

\appendix
\section{Diagonalization of $H_{\sf eff}$}\label{diagonalizza}
{\noindent}We show here how to diagonalize $H_{\sf eff} $ and find the polaritonic operators and frequencies. The results for $H$ simply follow by setting $\eta=u=0$. First, we define the (self-adjoint) canonical operators
\begin{align}
	\hat x_a&=\frac1{\sqrt{2\omega_a}}(\hat a+\!\hat a^\dagger),\quad \!\hat y_a=-i\sqrt{\frac{\omega_a}{2}}(\hat a-\hat a^\dagger)\\
	\hat x_b&=\frac1{\sqrt{2(\omega_b\!+\!4u)}}(\hat b+\hat b^\dagger),\quad\! \hat y_b\!=\!-i\sqrt{\frac{\omega_b\!+\!4u}{2}}(\hat b\!-\!\hat b^\dagger),
\end{align}
obeying the canonical commutation relations $[\hat x_i,\hat y_j]=i\delta_{ij}$. The Hamiltonian $H_{\sf eff}$, in the new coordinates $\hat{\mathbf{x}}=(\hat x_a,\hat x_b)$ and $\hat{\mathbf{y}}=(\hat y_a,\hat y_b)$ reads (up to a constant)
\begin{align}
	H_{\sf eff} &=\frac12 \hat{\mathbf{y}}^\intercal \hat{\mathbf{y}}+\frac12\hat{\mathbf{x}}^\intercal\mathbf{M}\hat{\mathbf{x}},\\
	\mathbf{M}&=\left(\begin{array}{cc} \omega_a^2+4D\omega_a&2\lambda\sqrt{\omega_a(\omega_b+4u)}\\
	2\lambda\sqrt{\omega_a(\omega_b+4u)} &(\omega_b-4\eta)(\omega_b+4u)
\end{array}\right).\label{M}
\end{align}
The above expression makes it evident that the eigenvalues of the matrix $\mathbf{M}$ correspond to the squared polaritonic frequencies $\omega_U^2,\omega_L^2$. From the same observation, we can also note that $\omega_U^2\omega_L^2={\sf det}\,\mathbf{M}$, which can be used to derive Eqs.~\eqref{product} and \eqref{prod2} of the main text in a simple way. We can diagonalize $\mathbf{M}$ via a $2\times2$ rotation matrix:
\begin{align}
	\mathbf{R}\mathbf{M}\mathbf{R}^\intercal={\sf diag}(\omega_U^2,\omega_L^2).\label{diagform}
\end{align}
One convenient way to find $\mathbf{R}$ is to expand $\mathbf{M}$ in the Pauli basis
\begin{align}
\mathbf{M}=\frac{1}{2}( M_0\mathbb{I}+ M_x \sigma_x + M_z\sigma_z),
\end{align}
where $M_0\!=\!{\sf Tr}(\mathbf{M}),M_j\!=\!{\sf Tr}(\mathbf{M}\sigma_j)$. Writing $\mathbf{R}\!=\!\bigl(\begin{smallmatrix}
\cos\theta&-\sin\theta\\ \sin\theta&\cos\theta
\end{smallmatrix} \bigr)={\rm e}^{-i\theta\sigma_y}$, we can see that
\begin{align}
\mathbf{R}\mathbf{M}\mathbf{R}^\intercal&=\frac{1}{2}[M_0\mathbb{I}+\sigma_x(M_x\cos2\theta+M_z\sin2\theta)\nonumber\\
&+\sigma_z(M_z\cos2\theta-M_x\sin2\theta)],
\end{align}
and the diagonalization corresponds to eliminating the $\sigma_x$ term. This can be achieved by choosing $\sin2\theta\!=\!-M_x/\sqrt{M_x^2+M_z^2},\cos2\theta\!=\! M_z/\sqrt{M_x^2+M_z^2}$, hence
\begin{align}
\mathbf{R}\mathbf{M}\mathbf{R}^\intercal &=\frac{1}{2}(M_0\mathbb{I}+\sqrt{M_x^2+M_z^2}\sigma_z).\label{diagonalM}
\end{align}
These calculations make it evident that the polaritonic frequencies are
\begin{align}
\omega_{U,L}^2=\frac{M_0\pm\sqrt{M_x^2+M_z^2}}{2},
\end{align}
which can be recast in terms of the model parameters if desired: this provides Eq.~\eqref{freqs} of the main text in the special case $\eta\!=\!0,u\!=\!0$. Note that our procedure is always consistent with the ordering of eigenvalues chosen in Eq.~\eqref{diagform}. To find the polaritonic annihilation operators, we can first define a new set of canonical operators as $(\hat X_U,\hat X_L)=\mathbf{R}\hat{\mathbf{x}}$ and $(\hat{Y}_U,\hat Y_L)=\mathbf{R}\hat{\mathbf{y}}$, such that
\begin{align}
	H &=\sum_{k=U,L}\frac12\left(\hat{Y}_k^2+\omega_k^2\hat{X}_k^2\right)+{\sf const.},
\end{align}
which is in the form of two decoupled harmonic oscillators with unit mass. The polaritonic annihilation operators can thus be written as
\begin{align}
	\hat p_k=\frac{1}{\sqrt2}\left(\sqrt{\omega_k}\hat X_k+i\frac{\hat Y_k}{\sqrt{\omega_k}}\right).
\end{align}
Performing the appropriate subsitutions, this gives
\begin{align}
	\!\!\!\hat p_{U}\!&=\!\cos\theta[\mu(\!\tfrac{\omega_U}{\omega_a}\!)\hat a\!+\!\nu(\!\tfrac{\omega_U}{\omega_a}\!)\hat a^\dagger]\!-\!\sin\theta[\mu(\!\tfrac{\omega_U}{\omega_b\!+4u})\hat b\!+\!\nu(\!\tfrac{\omega_U}{\omega_b\!+4u}\!)\hat b^\dagger],\label{pU1}\\
\!\!\hat p_{L}\!&=\!\sin\theta[\mu(\!\tfrac{\omega_L}{\omega_a}\!)\hat a\!+\!\nu(\!\tfrac{\omega_L}{\omega_a}\!)\hat a^\dagger]\!+\!\cos\theta[\mu(\!\tfrac{\omega_L}{\omega_b\!+4u}\!)\hat b\!+\!\nu(\!\tfrac{\omega_L}{\omega_b\!+4u}\!)\hat b^\dagger],\label{pL1}
\end{align}
where the explicit definition of the angle $\theta$ in terms of model parameters is given by
\begin{align}
\cos2\theta\!&=\!\frac{\omega_a^2+4D\omega_a\!-\!(\omega_b\!-\!4\eta)(\omega_b+4u)}{\omega_U^2\!-\!\omega_L^2}\label{cosenonew},
\end{align}
together with the condition $\theta<0$. 
\section{Analytical study of the populations}\label{popstudy}
{\noindent}Here we show in detail how the parameter $D$ impacts the relationship between $n_U$ and $n_L$ for the special case $\eta=u=0$. For convenience, we recall the expressions
\begin{align}
	n_U&=\cos^2\theta\,\nu(\tfrac{\omega_U}{\omega_a})^2+\sin^2\theta\,\nu(\tfrac{\omega_U}{\omega_b})^2 \label{nUa},\\
	n_L&=\sin^2\theta\,\nu(\tfrac{\omega_a}{\omega_L})^2+\cos^2\theta\,\nu(\tfrac{\omega_b}{\omega_L})^2\label{nLa},\\
	\omega_U\omega_L&=\omega_a\omega_b\sqrt{1+\tfrac{4}{\omega_a}(D-\tfrac{\lambda^2}{\omega_b})}.\label{producta}
\end{align}
We start with $D\!<\!\lambda^2/\omega_b$. This implies $\omega_U\omega_L\!<\!\omega_a\omega_b$; since $\omega_U\geq\max(\omega_a,\omega_b)$ [see Eq.~\eqref{freqs} in the main text], it must be also $\omega_L\!<\!\min(\omega_a,\omega_b)$. Therefore, one has $1\!\leq\!\tfrac{\omega_U}{\omega_a}\!<\! \tfrac{\omega_b}{\omega_L}$ and $1\!\leq\!\tfrac{\omega_U}{\omega_b}\!<\!\tfrac{\omega_a}{\omega_L}$. As $\nu(x)^2=\tfrac14(x+\tfrac{1}{x})-\tfrac12$ is monotonically increasing for $x\!\geq\!1$, we thus have $\nu(\tfrac{\omega_U}{\omega_a})^2\!<\!\nu(\tfrac{\omega_b}{\omega_L})^2$ and $\nu(\tfrac{\omega_U}{\omega_b})^2\!<\!\nu(\tfrac{\omega_a}{\omega_L})^2$, which yields $n_U\!<\! n_L$ [see Eqs.~(\ref{nUa},\ref{nLa})]. In the main text we have seen how $D\!=\!\lambda^2/\omega_b$ implies $n_U=n_L$. For $D>\lambda^2/\omega_b$ we resort to a direct calculation of the population difference. In Eqs.~(\ref{nUa},\ref{nLa}), we express $\sin^2\theta,\cos^2\theta$ in terms of $\cos2\theta$, for which we have an explicit expression. Performing tedious manipulations, we arrive at the result
\begin{align}
n_U\!-\! n_L\!&=\!\frac{(\omega_U\omega_L\!-\!\omega_a\omega_b)}{4\omega_a\omega_b\omega_U\omega_L(\omega_U\!-\!\omega_L)}F(D),\label{ndiff}\\
F(D)&\equiv4D\omega_a\omega_b\!-\!(\omega_U\omega_L\!-\!\omega_a\omega_b)(\omega_a\!+\!\omega_b),\label{FD}
\end{align}
where we note that the fraction in Eq.~\eqref{ndiff} is always positive due to $\omega_U>\omega_L$ and Eq.~\eqref{producta}, hence the sign is solely determined by $F(D)$. We note that $F(D)>0$ for $D=\lambda^2/\omega_b$, thus we ask whether $F$ can have a zero past this value. $F=0$ yields the two solutions
\begin{align}
	D_\pm\!=\!\frac{\omega_a\!+\!\omega_b}{8\omega_a}[(\omega_b\!-\!\omega_a)\!\pm\!\sqrt{(\omega_b\!-\!\omega_a)^2\!-\!16\omega_a\lambda^2/\omega_b}].
\end{align}
Note that $\omega_a\geq \omega_b$ always results in $D_\pm<0$ (or complex solutions when the argument of the square root is negative), hence $F$ never changes sign for $D>0$ and $n_U\!>\!n_L$. For $\omega_a<\omega_b$, we can have $D_{\pm}>0$ iff the square root is real, which requires $\lambda^2\leq\lambda_{\sf max}^2=\omega_b(\omega_b-\omega_a)^2/16\omega_a$. Assuming this to be the case, we focus on the smallest root of $F$. We define $D_{\sf max}\equiv D_-$, and note that the relation $x-\sqrt{x^2-y^2}\geq y^2/2x$, valid for $x>y>0$, can be used to bound
\begin{align}
	D_{\sf max}\geq\frac{\omega_b+\omega_a}{\omega_b-\omega_a}\frac{\lambda^2}{\omega_b},
\end{align}
where we can see that $D_{\sf max}$ is well above the TRK value when the bare frequencies of the two modes are of the same order. Hence, when $D$ reaches $D_{\sf max}$ one has again $n_U\!=\! n_L$. The results can be summarized as follows. The sign of $n_U-n_L$ is the same as that of $D-\lambda^2/\omega_b$ if:
\begin{align}
{\sf (i)}&\quad \omega_b\leq\omega_a,\\
{\sf (ii)}&\quad \omega_b>\omega_a\textrm{ and } \lambda\geq\lambda_{\sf max},\\
{\sf (iii)}&\quad \omega_b>\omega_a\textrm{ , } \lambda<\lambda_{\sf max}\textrm{ and }D\!<\!D_{\sf max}.
\end{align}
\section{Approximating matter as a collection of oscillators}\label{appb}
{\noindent}In the main text we have assumed that $H_{\sf mic}^0$ describes with good approximation a harmonically oscillating polarization field. To investigate inevitable deviations from this scenario,
we now assume that the bare matter Hamiltonian can be expanded as
\begin{align}
	H_{\sf mic}^0=\sum_{k}\omega_k\hat b_k^\dagger\hat b_k+\sum_n\Delta_n\hat\psi_n^\dagger\hat\psi_n,
\end{align}
where where the $\hat b_k$'s are mutually independent bosonic annihilation operators describing the harmonic excitations of interest (as in the main text), while $\psi_n=\ket 0\bra{\Delta_n}$ are annihilation operators for residual excited states $\ket{\Delta_n}$, of bare energy $\Delta_n$, that we aim to neglect in our problem. More specifically, we shall assume that the excited states $\ket{\Delta_n}$ are never significantly populated in the dynamics of interest, and can be adiabatically eliminated. We neglect the possibility of direct transitions between the excited states $\ket{\Delta_n}$ and the excited states of the $\hat b_k$'s, as they can be expected to provide higher order corrections in the relevant coupling constants. We assume that the current operator can be expanded linearly in terms of the fields of interest, according to
\begin{align}
\hat{\mathbf{j}}&=-\sum_{k}\boldsymbol{j}_k(\hat b_k+\hat b_k^\dagger)-\sum_n\boldsymbol{\cal J}_n(\hat \psi_n+\hat\psi_n^\dagger),	
\end{align}
where $\boldsymbol{\cal J}_n$ are appropriate constant vectors. Then
\begin{align}
&H_{\sf int}=\sum_k\boldsymbol{j}_k\!\cdot\!\hat{\mathbf{A}}(\hat b_k+\hat b_k^\dagger)+\sum_n\boldsymbol{\cal J}_n\!\cdot\!\hat{\mathbf{A}}(\hat \psi_n+\hat\psi_n^\dagger)+\kappa\hat A^2\label{genlin2},
\end{align}
where we introduced $\kappa=\sum_jq_j^2/2m_j$ for brevity. Consistency with the TRK sum rule implies that we can decompose
\begin{align}
\kappa\hat A^2=\sum_k\frac{(\boldsymbol{j}_k\!\cdot\!\hat{\mathbf{A}})^2}{\omega_k}+\sum_n\frac{(\boldsymbol{\cal J}_n\!\cdot\!\hat{\mathbf{A}})^2}{\Delta_n}.\label{A2pieces}
\end{align}
Note that, for our scopes, the second term on the right hand side of Eq.~\eqref{A2pieces} has to be much smaller than the first: this corresponds to the situation in which the modes $\hat b_k$ approximately saturate the allowed transitions from the bare ground state of matter. We shall follow the prescription of Schrieffer and Wolff, later generalized by Reiter and S\o rensen \cite{sorensen}, for the adiabatic elimination of excited subspaces. We decompose $H_{\sf mic}=H_g+H_e+V_++V_-$, where the support of $H_g$ is the low-energy subspace in which the dynamics of interest occurs (ground subspace for brevity), the support of $H_e$ is the weakly excited subspace to be eliminated (excited subspace for brevity), $V_-$ is the operator that induces transitions from the excited to the ground subspace, and $V_+=V_-^\dagger$. In our specific case one has 
\begin{align}
	H_g&\!=\!\sum_{k}\left[\omega_k\hat b_k^\dagger \hat b_k\!+\!\boldsymbol{j}_k\!\cdot\!\hat{\mathbf{A}}(\hat b_k\!+\!\hat b_k^\dagger)\!+\!\frac{(\boldsymbol{j}_k\!\cdot\!\hat{\mathbf{A}})^2}{\omega_k}\right]\!+\!\sum_n\!\!\frac{(\boldsymbol{\cal J}_n\!\cdot\!\hat{\mathbf{A}})^2}{\Delta_n},\\
	H_e&=\sum_n\Delta_n\hat\psi_n^\dagger\hat\psi_n,\\
	V_-&=(V_+)^\dagger=\sum_n\boldsymbol{\cal J}_n\!\cdot\!\hat{\mathbf{A}}\hat \psi_n
\end{align}
The adiabatic elimination of the excited subspace results in the effective Hamiltonian $H_{\sf eff}=H_g-V_-(H_e)^{-1}V_+$ \cite{sorensen}, where explicit calculation yields
\begin{align}
	V_-(H_e)^{-1}V_+=\sum_{n}\frac{(\boldsymbol{\cal J}_n\!\cdot\!\hat{\boldsymbol A})^2}{\Delta_n},\label{eff}
\end{align}
hence we have simply
\begin{align}
	H_{\sf eff}&=\sum_{k}\left[\omega_k\hat b_k^\dagger \hat b_k+\boldsymbol{j}_k\!\cdot\!\hat{\mathbf{A}}(\hat b_k+\hat b_k^\dagger)+\frac{(\boldsymbol{j}_k\!\cdot\!\hat{\mathbf{A}})^2}{\omega_k}\right].
\end{align}
Thus we have the important result that {\it the contribution of the neglected excitations should be removed from the $\hat A^2$ term}, and the remaining parameters are still related by a TRK-like rule. We note that the same technique can be easily applied to eliminate the high-frequency cavity and matter modes, as done in the main text, and that our calculation can be generalized to include the electrostatic term $V_{\sf img}$. To deal with the former case, we can simply truncate the bosonic modes to be eliminated to a single excitation, and treat them on the same footing as the modes $\psi_n$ (the error so introduced will be negligible if these modes remain close to their ground state). To deal with $V_{\sf img}$, we can write the matter dipole moment as
\begin{align}
\hat{\mathbf{d}}=&\sum_{k}i\frac{\boldsymbol{j}_k}{\omega_k}(\hat b_k-\hat b_k^\dagger)+i\sum_n\frac{\boldsymbol{\cal J}_n}{\Delta_n}(\hat \psi_n-\hat\psi_n^\dagger),
\end{align}
where we have maintained consistency with the commutation rule \eqref{Hdj}. Assuming a general bilinear form $V_{\sf img}=\tfrac12\sum_{ij}c_{ij}\hat d_i\hat d_j$ (with $c_{ij}$ appropriate constants depending on the cavity structure), we can express $V_{\sf img}$ in terms of the various matter operators, and identify new contributions to $H_g,H_e,V_+,V_-$. After having performed the adiabatic elimination (and truncated all expressions to second order in the coupling), we find that the only relevant contribution is
\begin{align}
V_{\sf img}&\simeq\frac12\sum_{kk'}\frac{\boldsymbol{j}_k\cdot\mathbf{c}\boldsymbol{j}_{k'}}{\omega_k\omega_{k'}}[i(\hat b_k\!-\!\hat b_k^\dagger)][i(\hat b_{k'}\!-\!\hat b_{k'}^\dagger)].
\end{align}
When only one mode is involved in the sum, it is easy to recast this expression as $V_{\sf img}\simeq u[i(\hat b-\hat b^\dagger)]^2$, $u\equiv{\boldsymbol{j}_1\cdot\mathbf{c}\boldsymbol{j}_{1}}/{\omega_b^2}$
\end{document}